\documentclass{article}
\usepackage{spconf,amsmath,graphicx,hyperref}
\usepackage{amssymb}
\usepackage{comment}
\usepackage{multirow}
\usepackage{multicol}
\usepackage{tabularx}
\usepackage{bbm}

\title{Mitigating data replication in text-to-audio generative diffusion models through anti-memorization guidance}
\name{Francisco Messina, Francesca Ronchini, Luca Comanducci, Paolo Bestagini, Fabio Antonacci\thanks{© 2025 IEEE. Personal use of this material is permitted. Permission from IEEE must be obtained for all other uses, in any current or future media, including reprinting/republishing this material for advertising or promotional purposes, creating new collective works, for resale or redistribution to servers or lists, or reuse of any copyrighted component of this work in other works 
The authors acknowledge support from the IEEE Signal Processing Society under the Signal Processing Society Scholarship Program. This work was partially supported by the European Union under the Italian National Recovery and Resilience Plan (NRRP) of NextGenerationEU (PE00000014 - program ``SERICS'').
}}
\address{Dipartimento di Elettronica, Informazione e Bioingegneria, Politecnico di Milano, Italy}
\begin{document}
\ninept
\maketitle
\begin{abstract}
A persistent challenge in generative audio models is data replication, where the model unintentionally generates parts of its training data during inference. In this work, we address this issue in text-to-audio diffusion models by exploring the use of anti-memorization strategies. We adopt Anti-Memorization Guidance (AMG), a technique that modifies the sampling process of pre-trained diffusion models to discourage memorization. Our study explores three types of guidance within AMG, each designed to reduce replication while preserving generation quality. We use Stable Audio Open as our backbone, leveraging its fully open-source architecture and training dataset. Our comprehensive experimental analysis suggests that AMG significantly mitigates memorization in diffusion-based text-to-audio generation without compromising audio fidelity or semantic alignment. 

\end{abstract}
\begin{keywords}
generative audio models, anti-memorization, copyright, generative AI, text-to-audio, text-to-music
\end{keywords}
\section{Introduction}
\label{sec:intro}
Automatic audio generation has grown rapidly with the rise of multimodal text-to-audio models, which generate short audio clips from user-provided captions~\cite{agostinelli2023musiclm,copet2023simple,liu2024audioldm}. Their accessibility has lowered technical barriers~\cite{ronchini2025paguri}, and their commercial success in music generation has raised ethical concerns, especially around copyright and intellectual property~\cite{passoni2025diffused,sturm2019artificial}. Diffusion-based techniques~\cite{ho2020denoising} have rapidly become one of the most widely adopted families of generative models. However, these models are also known to exhibit memorization, i.e., reproducing content directly from the training set, creating further concerns for intellectual property rights of the training data~\cite{bralios2024generation,carlini2023extracting}. This phenomenon has been studied extensively in the context of image generation, where duplicated images or captions in the training corpus, as well as highly specific user prompts, have been identified as primary drivers of memorization~\cite{somepalli2023diffusion,somepalli2023understanding,chen2024towards}. To address replication in diffusion models for image generation, several strategies have been explored~\cite{wen2024detecting}. One approach removes duplicated samples from the training set~\cite{carlini2023extracting}; while effective, it requires costly retraining. Prompt-based methods have also been investigated: randomization can reduce replication but often harms image fidelity~\cite{somepalli2023understanding}, whereas duplication suppresses certain concepts but demands substantial manual effort~\cite{kumari2023ablating}. In the audio domain, research on memorization in audio generative models remains limited. Bralios et al.~\cite{bralios2024generation} conduct an initial study on data replication in the TANGO model~\cite{ghosal2023text}, while Batlle et al.~\cite{batlle2024towards} evaluate metrics for their effectiveness in detecting memorization.

This paper presents the first study, to the best of our knowledge, on mitigating data replication in generative diffusion models for audio. Our approach builds on the Anti-Memorization Guidance (AMG) framework~\cite{chen2024towards}, originally proposed for image generation, which operates at inference time by steering the model away from memorized samples. We focus on two primary sources of memorization: highly detailed user prompts and repeated content in the training data~\cite{chen2024towards}. For our experiments, we use the Stable Audio Open model~\cite{evans2025stable}, a latent diffusion transformer~\cite{peebles2023scalable} capable of generating variable-length stereo audio. Both the model and its training dataset are fully open source, providing full transparency and control over the data used in our evaluation. The results show that AMG effectively mitigates memorization of training data, while preserving high audio quality. This confirms its ability to reduce replication without compromising the generative capabilities of the model.

Audio excerpts and the code used for the experiments are freely available online\footnote{\footnotesize \url{https://polimi-ispl.github.io/anti-memorization-tta/}}.

\section{Background and Preliminaries}
\label{sec:background}

This section provides a concise overview of the background necessary to facilitate a clear understanding of the study. 

\subsection{Latent Diffusion Models}
Let us define $\mathbf{x} \in \mathbb{R}^N$ as a discrete audio signal of length N samples. We consider an encoder $\mathcal{E}$ and a decoder $\mathcal{D}$, such that we can obtain the latent representation $\mathbf{z}= \mathcal{E}(\mathbf{x})$, from which we then retrieve the audio as $\mathbf{x} = \mathcal{D}(\mathbf{z})$. The latent diffusion model~\cite{rombach2022high}, is then characterized by a forward process, which gradually adds Gaussian noise to a latent vector $\mathbf{z}_0 \sim q(\mathbf{z})$ using $T$ timesteps, such that $\mathbf{z}_T\sim \mathcal{N}(0, \mathbf{I})$. Then it is possible to sample $\mathbf{z}_t$ at an arbitrary timestep $t$ as 
\begin{equation}
    q(\mathbf{z}_t|\mathbf{z}_0) = \mathcal{N}(\mathbf{z}_t, \sqrt{\overline{\alpha}_t} \mathbf{z}_0, (1-\overline{\alpha}_t \mathbf{I})),
\end{equation}
where $\overline{\alpha}_t = \prod_{s=0}^{t-1} (1-\beta_s)$ and $ \beta_t \in \mathbb{R}, \; t=0,\ldots, T-1$ is the noise schedule controlling the amount of noise added at each timestep. 
Then, in the backward process, at an arbitrary step $t$ the noise $\boldsymbol{\epsilon}_t$ is predicted via network $\epsilon_\theta$ by minimizing the loss
\begin{equation}
    \mathcal{L} = \mathbb{E}_{\boldsymbol{\epsilon}\sim \mathcal{N}(0, \mathbf{I}),t} [||\boldsymbol{\epsilon}_t - \epsilon_\theta(\mathbf{z}_t,t) ||_{2}^{2}],
\end{equation}counter-intuitive
where the conditioning can be simply inserted by adding a label to the argument of $\epsilon_\theta(\cdot)$.

\subsection{Classifier-Free Guidance}
Classifier-Free Guidance (CFG) is a technique that enables efficient conditioning by training at the same time over conditional and unconditional objectives~\cite{ho2021classifierfree}, by simply using an empty label during the unconditional case. During inference, the prediction is steered towards the conditional one using
\begin{equation}
    \hat{\boldsymbol{\epsilon}} \longleftarrow \epsilon_\theta(\mathbf{z}_t) + s_0 (\epsilon_\theta(\mathbf{z_t},y) - \epsilon_\theta(\mathbf{z}_t)),
    \label{eq:eps_hat}
\end{equation}
where $s_0>1$ controls the adjustment level and $y$ is the conditioning. Please note that the inference mechanism of CFG operates in a way that is conceptually similar to the mitigation strategies discussed in the remainder of this paper.

\begin{figure*}[htb]
\begin{minipage}[t]{.33\linewidth}
  \centering
  \includegraphics[width=\textwidth]{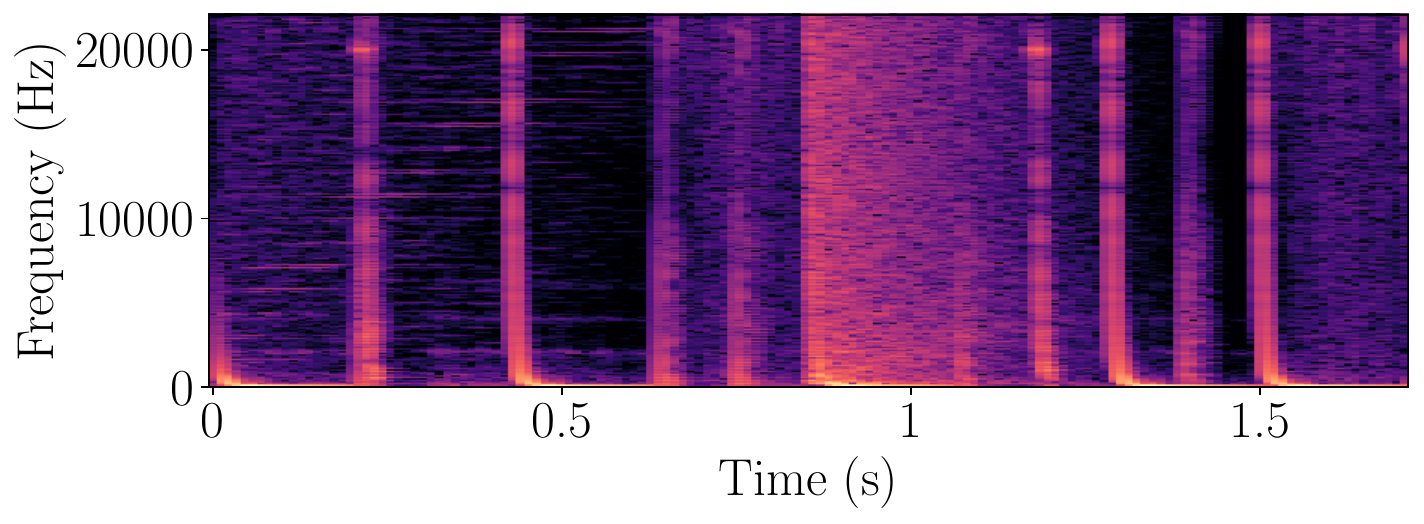}
  \centerline{(a)}
\end{minipage}
\hfill
\begin{minipage}[t]{.33\linewidth}
  \centering
  \includegraphics[width=\textwidth]{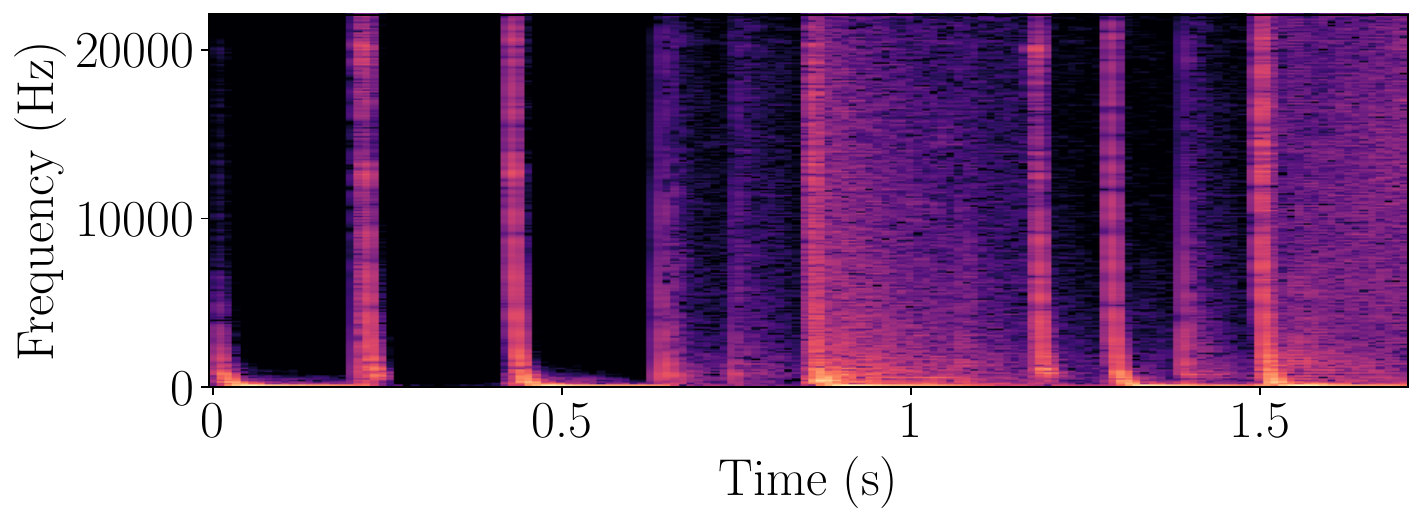}
  \centerline{(b)}
\end{minipage}
\hfill
\begin{minipage}[t]{.33\linewidth}
  \centering
  \includegraphics[width=\textwidth]{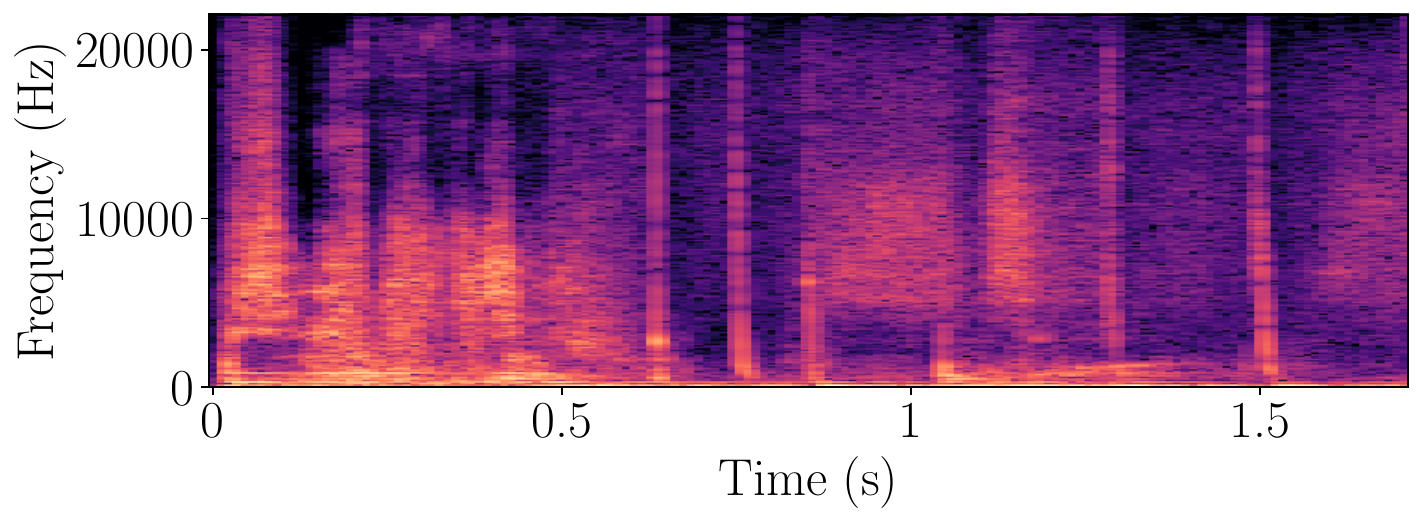}
  \centerline{(c)}\medskip
\end{minipage}
\caption{Example spectrograms of original training audio track (a) and generated using the same textual prompt without (b) and with AMG (c).}
\label{fig:spec_comparison}
\end{figure*}

\section{Method}
\label{sec:method}
This section introduces the Anti-Memorization Guidance (AMG) framework~\cite{chen2024towards}, adapted for audio latent diffusion models.

\subsection{Anti-memorization guidance framework}

The AMG framework continuously monitors the generation process, intervening only when the model’s output closely matches a known training example. 
To formalize this, let $\hat{\mathbf{x}} \in \mathbb{R}^N$ denote audio generated by a diffusion model, and let $\mathbf{x} \in \mathbb{R}^N$ represent an audio sample from the training dataset $\mathcal{X}$. The most similar training example can be identified by finding its nearest neighbor $\boldsymbol{\nu}$, computed as the sample in $\mathcal{X}$ with the smallest $\ell_2$ distance to $\hat{\mathbf{x}}$.

Since computing the  $\ell_2$ norm between the 1D waveforms is highly sensitive to time shifts and other variations, we instead compute the distance between descriptors extracted via a network.   
Specifically, CLAP\textsubscript{\textit{laion}}~\cite{laionclap2023}, following the approach of~\cite{bralios2024generation}, and denote the CLAP embedding function as $f_\text{CLAP}(\cdot)$. The nearest neighbor in descriptor space is then defined as:  
\begin{equation}
    \boldsymbol{\nu} = \arg\min_{\mathbf{x} \in \mathcal{X}} \; \| f_\text{CLAP}(\hat{\mathbf{x}}) - f_\text{CLAP}(\mathbf{x}) \|_2.
    \label{eq:nearest_neighbor}
\end{equation}

We then define a similarity measure to quantify the degree of memorization for an individual track, based on the cosine distance between the generated track and its nearest neighbor: 
\begin{equation}
    \sigma_t = \frac{f_\text{CLAP}(\hat{\mathbf{x}}) \cdot f_\text{CLAP}(\mathbf{x})}{\lVert f_\text{CLAP}(\hat{\mathbf{x}}) \rVert \, \lVert f_\text{CLAP}(\mathbf{x}) \rVert}.
    \label{eq:similarity}
\end{equation}
The AMG framework introduces three complementary guidance strategies, each targeting a distinct source of memorization, in the form of vectors summed to the noise in order to steer the prediction away from the memorized data. Specifically, it includes: \textit{despecification guidance} $\mathbf{g}_{\text{spe}}$, which mitigates the influence of the user’s prompt; \textit{caption deduplication guidance} $\mathbf{g}_{\text{dup}}$, which uses a memorized example’s caption as a negative prompt; and \textit{dissimilarity guidance} $\mathbf{g}_{\text{sim}}$, which actively steers the generation away from its nearest neighbor in the embedding space. AMG operates exclusively at inference time, applied directly during the reverse process. Concretely, the noise $\boldsymbol{\epsilon} \sim \mathcal{N}(\mathbf{0}, \mathbf{I})$ is updated at each inference step as: 
\begin{equation}
    \hat{\boldsymbol{\epsilon}} \longleftarrow \hat{\boldsymbol{\epsilon}} + \mathbbm{1}_{\{\sigma_t > \lambda_t\}}\bigl(\mathbf{g}_{\text{spe}} + \mathbf{g}_{\text{dup}} + \mathbf{g}_{\text{sim}}\bigr),
    \label{eq:eps_hat}
\end{equation}
where $\lambda_t$ is a pre-selected threshold and $\mathbbm{1}_{(\cdot)}$ an indicator function that actuates guidance when the current similarity level exceeds the threshold. Then, the estimate of the latent vector at timestep $t-1$ during the backward process can be computed as: 
\begin{equation}
    \mathbf{z}_{t-1} \longleftarrow \sqrt{\bar{\alpha}_{t-1}}\Bigl(\tfrac{\mathbf{z}_t - \sqrt{1 - \bar{\alpha}_t}\,\boldsymbol{\hat\epsilon}}{\sqrt{\bar{\alpha}_t}}\Bigr)+\sqrt{1 - \bar{\alpha}_{t-1}}\boldsymbol{\hat\epsilon}.
\end{equation}


\subsection{Despecification guidance}
\textit{Despecification guidance} mitigates memorization caused by overly specific prompts, which can act as ``keys''
leading to the generation of audio identical to examples in the training set. Given the noised latent space vector $\mathbf{z}_t$, we can obtain the corresponding prediction $\hat{\mathbf{z}}_0$ using the diffusion kernel
\begin{equation}
    \hat{\mathbf{z}}_0 = \frac{\mathbf{z}_t -\sqrt{1 - \bar\alpha_t}\hat{\boldsymbol{\epsilon}}}{\sqrt{\bar\alpha_t}}.
    \label{eq:z_0_hat}
\end{equation}
Then, we estimate the audio signal $\hat{\mathbf{x}} = \mathcal{D}(\hat{\mathbf{z}}_0)$, find its nearest neighbor $\boldsymbol{\nu}$ and compute a similarity metric $\sigma_t(\hat{\mathbf{x}}, \boldsymbol{\nu})$ as defined in \eqref{eq:nearest_neighbor} and \eqref{eq:similarity}, respectively. The despecification guidance operates oppositely to CFG, specifically by reducing the influence of the conditioning on the predicted noise,  which is achieved by computing the vector

\begin{equation}
    \mathbf{g}_{\mathrm{spe}} = -s_1 \bigl(\mathbf{\epsilon}_\theta(\mathbf{z}_t, y) - \epsilon_\theta(\mathbf{z}_t)\bigr),
\end{equation}
where $\epsilon_\theta(\mathbf{z}_t,y)$ represent the prediction conditioned on the text prompt, and $\epsilon_\theta(\mathbf{z}_t)$ corresponds to the unconditional prediction. The guidance scale $s_1$ is defined as
\begin{equation}
    s_1 = \max\bigl(\min(c_1\,\sigma_t, s_0 - 1), 0\bigr),
    \label{eq:despec_scale}
\end{equation}
where $c_1 \in \mathbb{R}$ is a constant and $s_0$ is the original guidance scale. With this definition, the new guidance scale $c_1 \sigma_t$, defined at step $t$, varies at each step according to the similarity of the reconstructed audio.  
The $\max(\cdot, 0)$ function ensures the scale remains non-negative,  while the $\min(\cdot)$ prevents excessive deviation from the input prompt.

\subsection{Caption Deduplication Guidance}
\textit{Caption deduplication guidance} mitigates memorization caused by duplicated captions in the training set. 
Specifically, it identifies such duplicated captions and steers the generation away from them.
The deduplication guidance vector is computed as
\begin{equation}
    \mathbf{g}_{\mathrm{dup}} = -s_{2} \bigl( \epsilon_{\theta}(\mathbf{z}_{t}, y_{\nu}) - \epsilon_{\theta}(\mathbf{z}_{t}) \bigr),
    \label{eq:dup_guidance}
\end{equation}
where $y_\nu$ is the caption corresponding to the nearest neighbor $\boldsymbol{\nu}$.  
The deduplication guidance scale is defined as
\begin{equation}
    s_{2} = \max \Bigl( \min \bigl( c_{2} \sigma_{t}, s_{0} - s_{1} - 1 \bigr), 0 \Bigr),
    \label{eq:dup_scale}
\end{equation}
where the $\min(\cdot)$ term accounts for the despecification guidance to properly bound the prediction, ensuring the generation is not excessively steered away from the input prompt.  
In cases of duplicated captions, the nearest neighbor $\boldsymbol{\nu}$ ideally corresponds to the memorized sample.

\subsection{Dissimilarity Guidance}
The \textit{Dissimilarity Guidance} differs from the other two by explicitly reducing the similarity score, achieved by minimizing its gradient with respect to the ground-truth prediction at time $t$.  
Specifically, we compute the vector
\begin{equation}
    \mathbf{g}_\text{sim} = c_3 \sqrt{1-\overline{\alpha}_t} \, \nabla_{x_t} \sigma_t,
\end{equation}
where $\nabla$ denotes the gradient and $c_3 \in \mathbb{R}$ is a parameter controlling the strength of the guidance.  
Minimizing the similarity gradient in this way effectively suppresses memorization.

\section{Results and discussion}

\label{sec:results}
This section presents results that show AMG’s effectiveness in reducing data memorization.

\subsection{Experimental setup}
We use Stable Audio Open~\cite{evans2025stable}, which provides publicly available checkpoints. Our dataset consists of 6,000 tracks from Stable Audio Open 1.0, primarily sourced from Freesound~\cite{fonseca2017freesound} and FMA~\cite{michael_defferrard_2017_1414728}.
It includes both music and foley sound examples, all sampled at $44100~\mathrm{Hz}$. To ensure that the selected samples were as diverse as possible, we clustered the dataset using k-nearest neighbor (k-NN) clustering on fused embeddings, obtained by merging CLAP\textsubscript{\textit{laion}}~\cite{laionclap2023} audio and text embeddings to capture both acoustic and semantic features.  
For each cluster, we chose the sample from its densest region, which is most likely to trigger memorization, resulting in $60$ examples. For each experiment, we used $100$ denoising steps, a CFG scale of $s_0 = 7$, a parabolic $\lambda_t$ schedule ranging from $0.4$ to $0.5$, and a new random seed for each generation.  
We report both qualitative and quantitative results, including similarity analysis and t-Distributed Stochastic Neighbor Embedding (t-SNE)~\cite{maaten2008visualizing} visualization, to evaluate the effectiveness of our approach. We also conducted ablation studies over all possible combinations of the AMG guidance strategies to isolate their individual and joint contributions to reducing memorization, using mean similarity, prompt adherence, and audio quality as evaluation metrics. The guidance coefficients were kept constant across all experimental conditions. Because despecification and caption deduplication interact through $s_1$ and $s_2$, we set both $c_1$ and $c_2$ equal to $s_0 - 1$, i.e., the CFG scale minus one.  For dissimilarity guidance, we set $c_3 = 1000$, based on empirical tuning. The baseline corresponds to the standard generation process without applying any mitigation strategies.

\subsection{Qualitative Analysis}

\begin{figure}[htb]
\begin{minipage}[b]{.49\columnwidth}
  \centering
\centerline{\includegraphics[width=\textwidth]{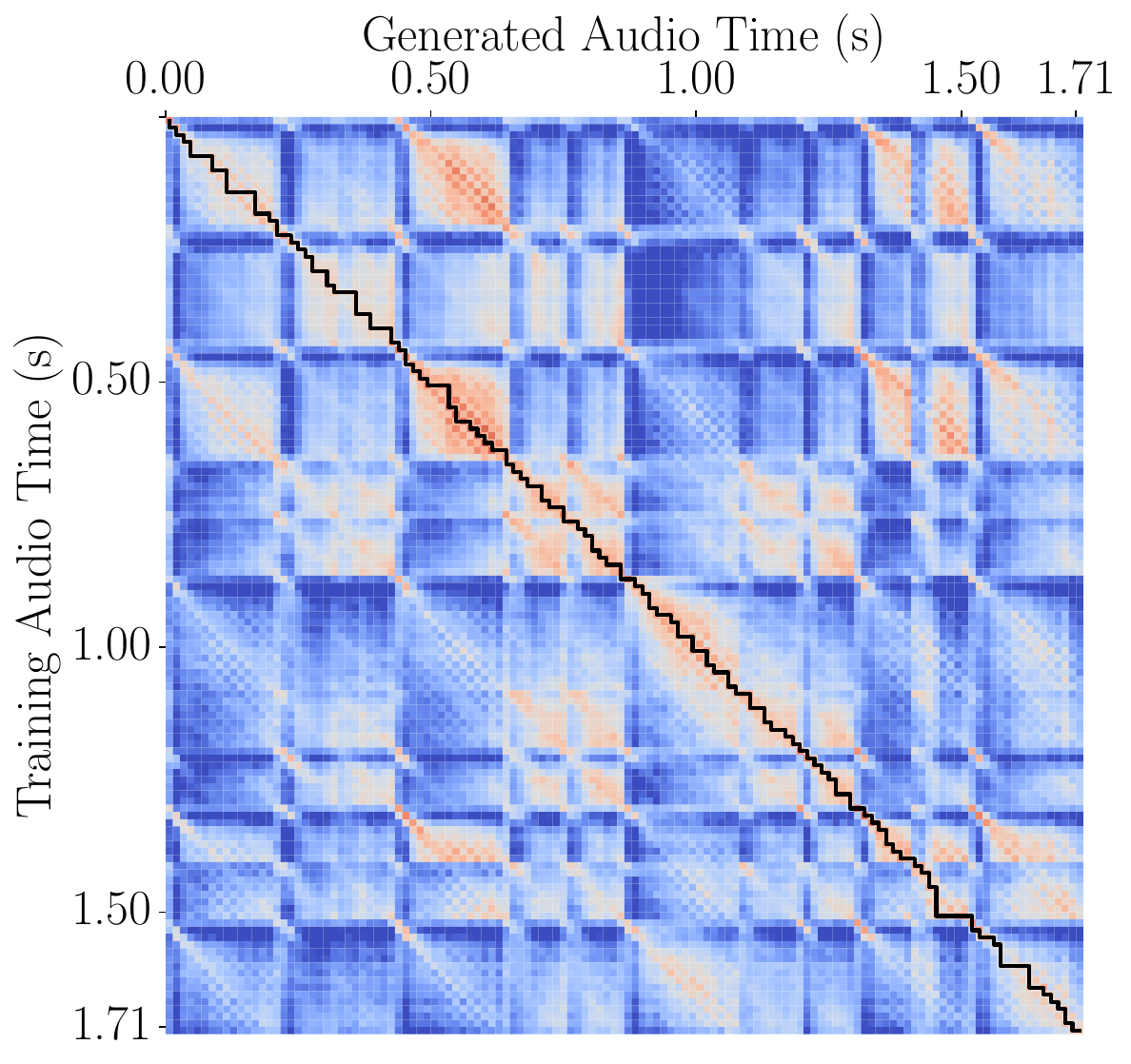}}
  \centerline{(a)}\medskip
\end{minipage}
\hfill
\begin{minipage}[b]{.49\columnwidth}
  \centering
\centerline{\includegraphics[width=\textwidth]{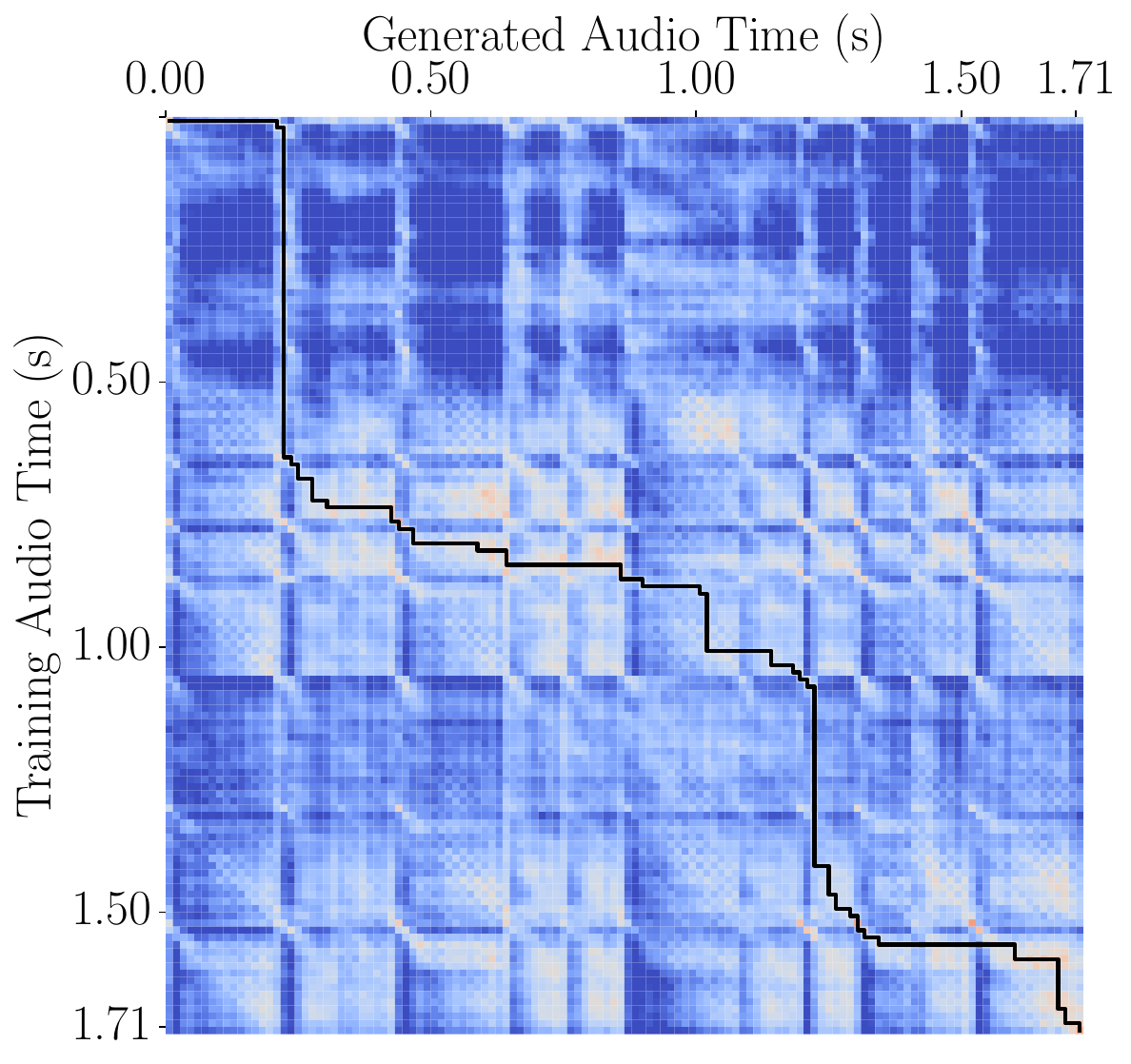}}
  \centerline{(b)}\medskip
\end{minipage}

\caption{Similarity matrices computed on the same audio track considered in Fig~\ref{fig:spec_comparison} without (a) and with AMG (b).}
\label{fig:similarity_matrices}
\end{figure}
\begin{figure}[htb]
\centering
\begin{minipage}[t]{.49\columnwidth}
  \centering
\includegraphics[width=\columnwidth]{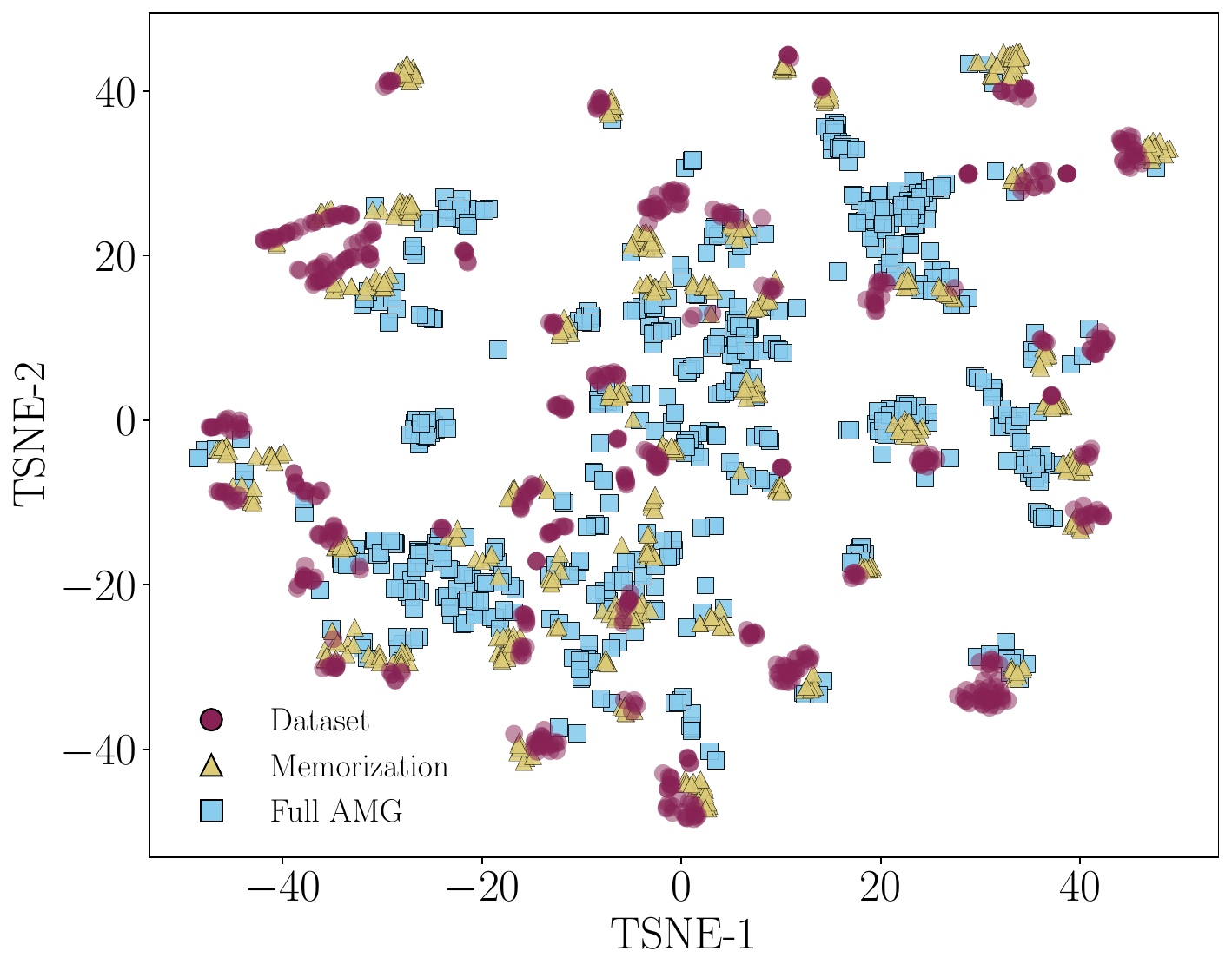} 
\centerline{(a)}
\medskip
\end{minipage}
\hfill
\begin{minipage}[t]{.49\columnwidth}
  \centering
\includegraphics[width=\columnwidth]{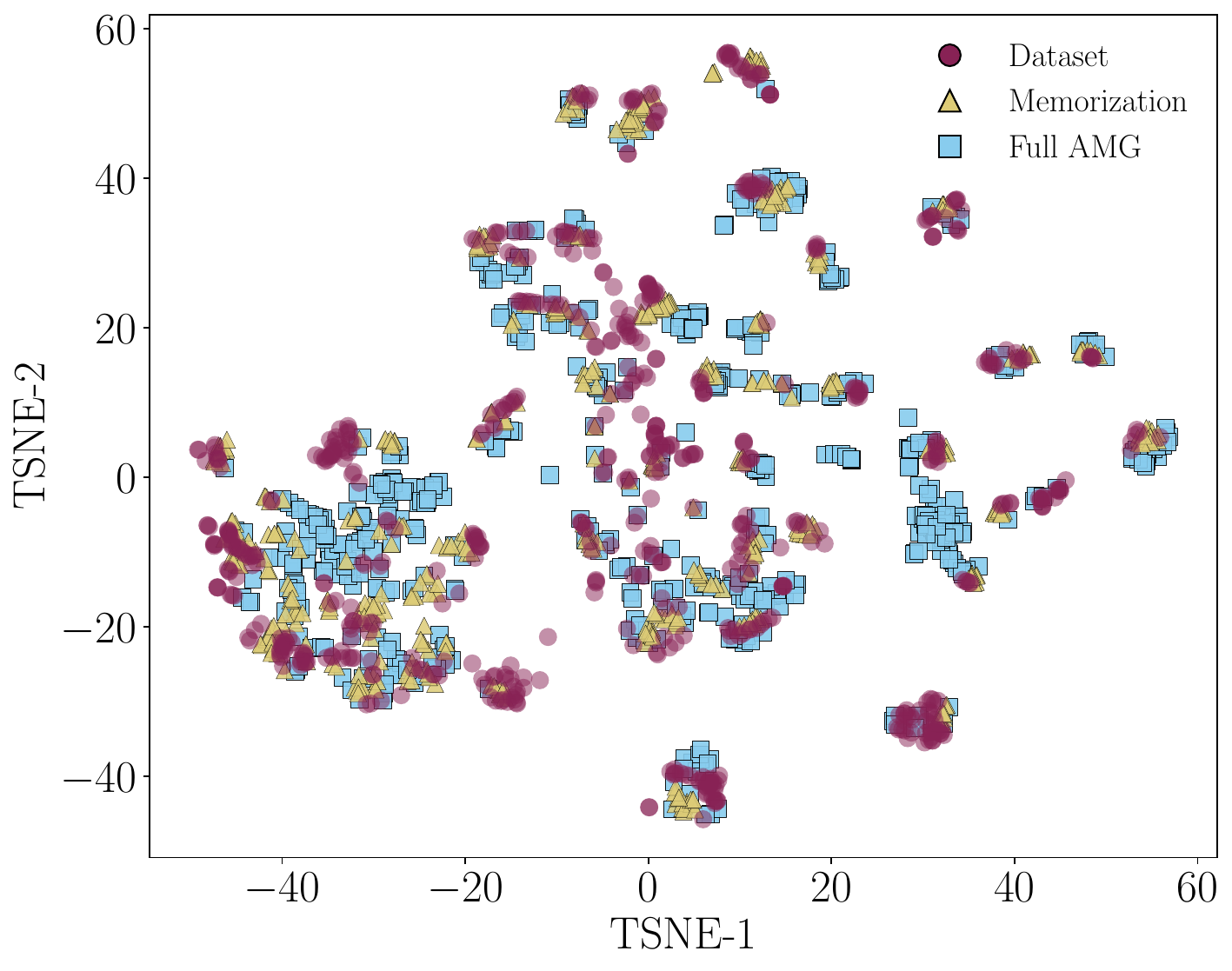}
  \centerline{(b)}\medskip
\end{minipage}
\caption{T-SNE visualization of embeddings from the considered dataset and from audio generated with (\textit{Full AMG}) and without AMG (\textit{Memorization}), using CLAP\textsubscript{\textit{laion}} (a) and MERT (b) as embedding extractors.}
\label{fig:tsne-2}
\end{figure}

\begin{figure}[htb]
\centering
\begin{minipage}[t]{.9\columnwidth}
  \centering
\includegraphics[width=\columnwidth]{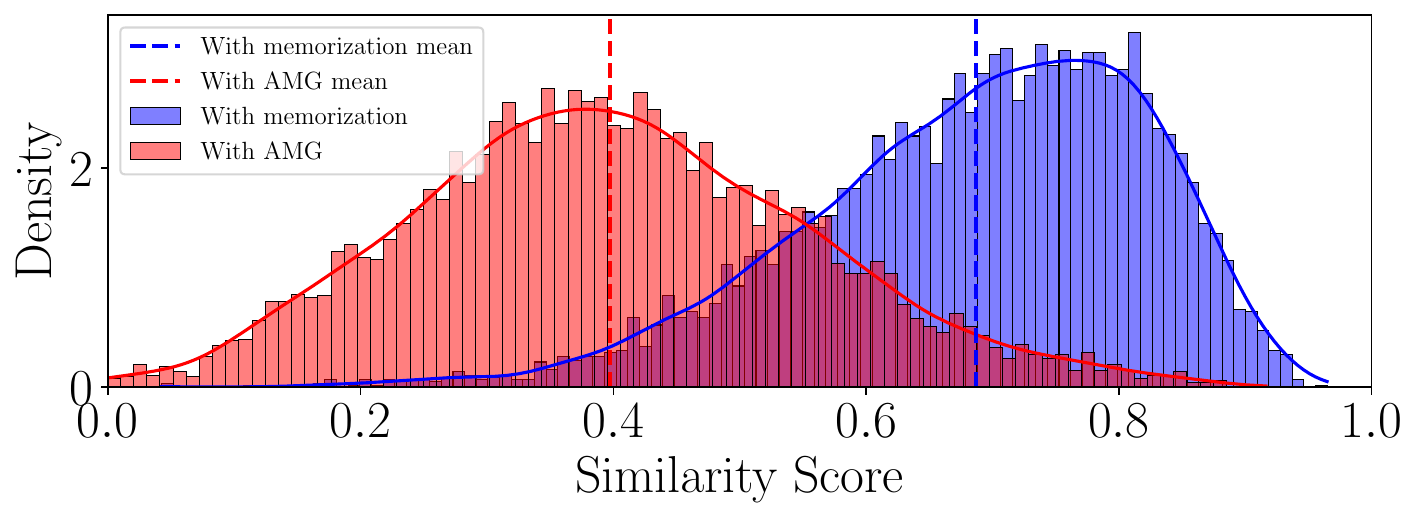} 
\centerline{(a)}
\end{minipage}
\begin{minipage}[t]{.9\columnwidth}
  \centering
\includegraphics[width=\columnwidth]{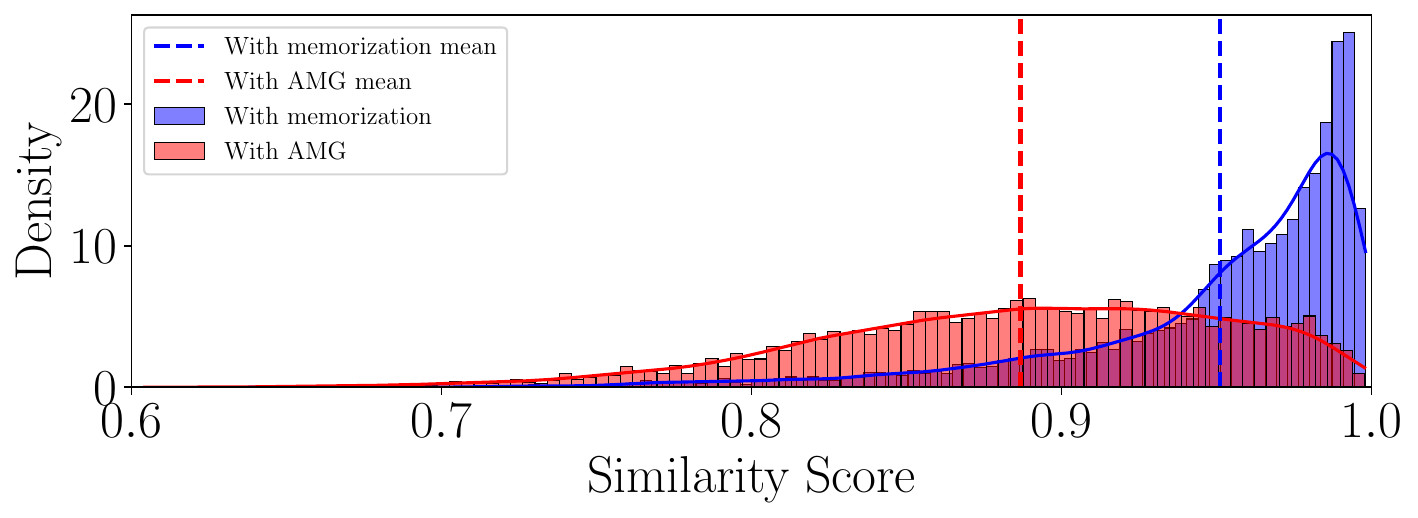}
  \centerline{(b)}\medskip
\end{minipage}
\vspace{-6pt}
\caption{Histogram of similarity score distributions computed over embeddings extracted via CLAP\textsubscript{\textit{laion}} (a) and MERT (b).}
\label{fig:histogram-2}
\end{figure}
\begin{table*}[hbt]
    \centering
    \caption{Ablation Study of Guidance Strategies. Columns report: mean similarity (CLAP\textsubscript{\textit{laion}} and MERT), prompt adherence (CLAPscore), Fréchet Audio Distance (FAD), Kernel Audio Distance (KAD), and Mauve Audio Divergence (MAD). }
    \label{tab:merged_results}
    \begin{tabular}
    {p{8em}|c|c|c|c|c|c|cc}
        \hline
        \multirow{2}{*}{\textbf{Guidance strategy}} 
        & \multicolumn{2}{c|}{\textbf{Mean Similarity} $~\downarrow$} 
        & \multicolumn{1}{c|}{\textbf{Prompt Adherence}$~\uparrow$} 
        & \multicolumn{2}{c|}{\textbf{FAD}$~\downarrow$} 
        & \multicolumn{1}{c|}{\textbf{KAD}$~\downarrow$} 
        & \multicolumn{1}{c}{\textbf{MAD}$~\downarrow$} \\
    \cline{2-3} \cline{4-4} \cline{5-6} \cline{7-7} \cline{8-8}
        & CLAP\textsubscript{\textit{laion}} & MERT & CLAP\textsubscript{\textit{laion}} & CLAP\textsubscript{\textit{laion}} & MERT & CLAP\textsubscript{\textit{laion}} & MERT \\
        \cline{4-4}
        \hline
        Baseline & 0.69 & 0.95 & \textbf{0.32} & 0.17 & 4.27 & 17.20 & 4.22 \\
        \hline
        $\mathbf{g}_{spe}$ & 0.69 & 0.95 & 0.32 & 0.16 & 4.15 & \textbf{16.91} & 3.65 \\
        $\mathbf{g}_{dup}$ & 0.64 & 0.93 & 0.27 & 0.16 & 3.94 & 17.13 & 4.36 \\
        $\mathbf{g}_{sim}$ & 0.41 & 0.90 & 0.20 & 0.15 & 3.55 & 18.24 & 3.50 \\
        $\mathbf{g}_{spe} + \mathbf{g}_{dup}$ & 0.62 & 0.93 & 0.25 & 0.15 & 3.29 & 17.67 & 3.10 \\
        $\mathbf{g}_{spe} + \mathbf{g}_{sim}$ & 0.43 & 0.90 & 0.19 & 0.15 & 2.93 & 17.61 & 3.12 \\
        $\mathbf{g}_{dup} + \mathbf{g}_{sim}$ & 0.41 & 0.89 & 0.16 & 0.15 & 2.84 & 17.72 & 3.11 \\
        \hline
        Full AMG (All) & \textbf{0.40} & \textbf{0.89} & 0.14 & \textbf{0.15} & \textbf{2.57} & 18.27 & \textbf{2.74} \\
        \hline
    \end{tabular}
\end{table*}

To visually show the impact of AMG, we present spectrograms computed from an audio sample\footnote{\url{https://freesound.org/people/Jovica/sounds/5131/}} in the dataset. Fig.~\ref{fig:spec_comparison}(a) shows the spectrogram of the original audio, while Fig.~\ref{fig:spec_comparison}(b) and \ref{fig:spec_comparison}(c) display spectrograms of audio generated with the same prompt, with and without AMG, respectively.  
The spectrograms show that the temporal and spectral evolution of audio generated with AMG diverges significantly from the training data, in contrast to the version produced without any mitigation strategy. To further highlight this difference, we examine the same excerpts using the structural similarity matrix~\cite{cooper2003summarizing}. 
This method segments a track into short overlapping windows and computes the cosine similarity between CLAP\textsubscript{\textit{laion}} embeddings of all segment pairs.  
Fig.~\ref{fig:similarity_matrices}(a) and (b) show the structural similarity matrices for the same audio example without and with AMG, respectively. Examining the black line, which highlights the highest similarity value in each row, we observe that without mitigation, these values are concentrated along the diagonal, indicating an almost one-to-one temporal correspondence between the original and generated audio. When AMG is applied, the highest similarity values shift away from the diagonal, demonstrating that the generated audio differs substantially from the training data. Fig.~\ref{fig:tsne-2} presents a t-SNE visualization of the dataset, along with samples generated with and without AMG. Visual inspection reveals that samples from the original dataset and those generated without mitigation cluster closely together. In contrast, samples produced with AMG are more dispersed and irregularly distributed, highlighting AMG’s effectiveness in reducing similarity and memorization.

\subsection{Similarity analysis}
We quantify similarity by representing each generated audio clip with a single embedding and computing the cosine distance between this embedding, its original counterpart, and the corresponding nearest neighbor in the training set. As embedding extractors we consider CLAP\textsubscript{\textit{laion}}~\cite{laionclap2023}, which focuses on the semantic characteristics of the audio, and MERT~\cite{li2024mert}, which is a self-supervised audio embedding model trained for acoustic music analysis. In Fig.~\ref{fig:histogram-2} we aggregate these similarity values across audio examples, with or without AMG. Fig.~\ref{fig:histogram-2} shows the distribution of similarity values computed across all audio examples, comparing cases with and without AMG. Fig.~\ref{fig:histogram-2}(a) reports results using CLAP\textsubscript{\textit{laion}} embeddings, while Fig.~\ref{fig:histogram-2}(b) reports results using MERT embeddings. The red histograms correspond to samples generated with AMG, and the blue histograms correspond to those generated without AMG. The dashed lines indicate the mean of the distributions, showing that the application of AMG consistently shifts the mean toward lower similarity values, thereby reducing memorization.
To comprehensively assess similarity, we perform an ablation study evaluating all combinations of AMG techniques; mean similarity values are reported in Table~\ref{tab:merged_results} (mean similarity columns). Dissimilarity guidance is the most effective individual component, substantially reducing similarity on its own. In contrast, despecification guidance has negligible impact, and caption-deduplication guidance yields only minor gains over the baseline. Combining strategies generally improves performance, with any configuration including $\mathbf{g}_{\mathrm{sim}}$ achieving the strongest reductions. The full AMG framework attains the lowest mean similarity, confirming the advantage of applying all strategies jointly.

\subsection{Prompt Adherence}
We measure prompt adherence using CLAPScore, defined as the cosine similarity between the CLAP audio embedding of each generated sample and the CLAP text embedding of its corresponding caption~\cite{majumder2024tango, liu2024audioldm}. Table~\ref{tab:merged_results} reports mean values across ablation conditions. Results reveal a trade-off between reducing memorization and preserving prompt fidelity: as the steering effect of AMG increases, similarity to training references decreases substantially, together with prompt adherence. This is expected, as guidance mechanisms discouraging memorization can inherently conflict with prompt alignment. These findings highlight the importance of AMG hyperparameters, which must be carefully tuned to balance the generation of novel content with faithfulness to user intent.

\subsection{Audio Quality}
We evaluate audio quality using Fréchet Audio Distance (FAD)~\cite{kilgour2018fr,gui2024adapting}, along with the newer Kernel Audio Distance (KAD)~\cite{kad} and Mauve Audio Divergence (MAD)~\cite{huang2025aligning}, which better correlate with human judgments. As reference, we use the 6,000 audio files in this study. FAD is computed using MERT~\cite{gui2024adapting} and CLAP\textsubscript{\textit{laion}} embeddings, KAD using CLAP\textsubscript{\textit{laion}}, and MAD using MERT. Results in Table~\ref{tab:merged_results} show a notable decrease in FAD for the full AMG condition across all embedding models. This outcome is counterintuitive, as steering interventions are often expected to reduce perceptual quality. A possible explanation is that the baseline’s memorization produces repetitive outputs that deviate from the diverse reference distribution, while AMG encourages greater variability, yielding generations that better align with the reference. Further investigation is needed to fully understand this effect.

\section{Conclusion}
\label{sec:conclusion}
This paper introduces anti-memorization guidance to mitigate data memorization in diffusion-based generative audio models by steering inference away from memorized content. Experiments on similarity, prompt adherence, and audio quality show that reducing memorization is both practical and effective, underscoring the potential of this approach and motivating further research. Future work will explore extending anti-memorization guidance to other modalities and integrating it with training-time interventions.


\bibliographystyle{IEEEbib}
\bibliography{refs,strings}

@inproceedings{chen2024towards,
  title={Towards memorization-free diffusion models},
  author={Chen, Chen and Liu, Daochang and Xu, Chang},
  booktitle={IEEE/CVF CVPR},
  year={2024}
}

@inproceedings{batlle2024towards,
  title={Towards assessing data replication in music generation with music similarity metrics on raw audio},
  author={Batlle-Roca, Roser and Liao, Wei-Hsiang and Serra, Xavier and Mitsufuji, Yuki and G{\'o}mez, Emilia},
  booktitle={ISMIR, San
Francisco, USA},
  year={2024}
}

@inproceedings{bralios2024generation,
  title={Generation or replication: Auscultating audio latent diffusion models},
  author={Bralios, Dimitrios and Wichern, Gordon and Germain, Fran{\c{c}}ois G and Pan, Zexu and Khurana, Sameer and Hori, Chiori and Le Roux, Jonathan},
  booktitle={ICASSP},
  year={2024},
  organization={IEEE}
}

@inproceedings{
li2024mert,
title={{MERT}: Acoustic Music Understanding Model with Large-Scale Self-supervised Training},
author={Yizhi LI and Ruibin Yuan and Ge Zhang and Yinghao Ma and Xingran Chen and Hanzhi Yin and Chenghao Xiao and Chenghua Lin and Anton Ragni and Emmanouil Benetos and others },
booktitle={The Twelfth ICLR},
year={2024},
}

@inproceedings{evans2025stable,
  title={Stable audio open},
  author={Evans, Zach and Parker, Julian D and Carr, CJ and Zukowski, Zack and Taylor, Josiah and Pons, Jordi},
  booktitle={ICASSP},
  year={2025},
  organization={IEEE}
}

@inproceedings{
ho2021classifierfree,
title={Classifier-Free Diffusion Guidance},
author={Jonathan Ho and Tim Salimans},
booktitle={NeurIPS 2021 Workshop on Deep Generative Models and Downstream Applications},
year={2021},
}

@inproceedings{cooper2003summarizing,
  title={Summarizing popular music via structural similarity analysis},
  author={Cooper, Matthew and Foote, Jonathan},
  booktitle={WASPAA},
  year={2003},
  organization={IEEE}
}

@article{maaten2008visualizing,
  title={Visualizing data using t-SNE},
  author={Maaten, Laurens van der and Hinton, Geoffrey},
  journal={JMLR},
  volume={9},
  number={Nov},
  pages={2579--2605},
  year={2008}
}

@article{kilgour2018fr,
  title={Fr\'{e}chet audio distance: A metric for evaluating music enhancement algorithms},
  author={Kilgour, Kevin and Zuluaga, Mauricio and Roblek, Dominik and Sharifi, Matthew},
  journal={arXiv preprint arXiv:1812.08466},
  year={2018}
}

@inproceedings{gui2024adapting,
  title={Adapting frechet audio distance for generative music evaluation},
  author={Gui, Azalea and Gamper, Hannes and Braun, Sebastian and Emmanouilidou, Dimitra},
  booktitle={ICASSP},
  year={2024},
  organization={IEEE}
}

@Article{ronchini2025paguri,
AUTHOR = {Ronchini, Francesca and Comanducci, Luca and Perego, Gabriele and Antonacci, Fabio},
TITLE = {{PAGURI}: A User Experience Study of Creative Interaction with Text-to-Music Models},
JOURNAL = {Electronics},
VOLUME = {14},
YEAR = {2025},
NUMBER = {17},
ARTICLE-NUMBER = {3379},
}

@inproceedings{peebles2023scalable,
  title={Scalable diffusion models with transformers},
  author={Peebles, William and Xie, Saining},
  booktitle={IEEE/CVF ICCV},
  year={2023}
}

@article{agostinelli2023musiclm,
  title={Musiclm: Generating music from text},
  author={Agostinelli, Andrea and Denk, Timo I and Borsos, Zal{\'a}n and Engel, Jesse and Verzetti, Mauro and Caillon, Antoine and Huang, Qingqing and Jansen, Aren and Roberts, Adam and Tagliasacchi, Marco and others},
  journal={arXiv preprint arXiv:2301.11325},
  year={2023}
}

@article{liu2024audioldm,
  title={Audioldm 2: Learning holistic audio generation with self-supervised pretraining},
  author={Liu, Haohe and Yuan, Yi and Liu, Xubo and Mei, Xinhao and Kong, Qiuqiang and Tian, Qiao and Wang, Yuping and Wang, Wenwu and Wang, Yuxuan and Plumbley, Mark D},
  journal={IEEE/ACM Trans. Audio Speech Lang. Process},
  volume={32},
  pages={2871--2883},
  year={2024},
  publisher={IEEE}
}

@article{copet2023simple,
  title={Simple and controllable music generation},
  author={Copet, Jade and Kreuk, Felix and Gat, Itai and Remez, Tal and Kant, David and Synnaeve, Gabriel and Adi, Yossi and D{\'e}fossez, Alexandre},
  journal={NeurIPS},
  volume={36},
  pages={47704--47720},
  year={2023}
}

@article{ho2020denoising,
  title={Denoising diffusion probabilistic models},
  author={Ho, Jonathan and Jain, Ajay and Abbeel, Pieter},
  journal={NeurIPS},
  volume={33},
  pages={6840--6851},
  year={2020}
}

@inproceedings{carlini2023extracting,
  title={Extracting training data from diffusion models},
  author={Carlini, Nicolas and Hayes, Jamie and Nasr, Milad and Jagielski, Matthew and Sehwag, Vikash and Tramer, Florian and Balle, Borja and Ippolito, Daphne and Wallace, Eric},
  booktitle={32nd USENIX security symposium},
  year={2023}
}

@inproceedings{somepalli2023diffusion,
  title={Diffusion art or digital forgery? investigating data replication in diffusion models},
  author={Somepalli, Gowthami and Singla, Vasu and Goldblum, Micah and Geiping, Jonas and Goldstein, Tom},
  booktitle={IEEE/CVF CVPR},
  year={2023}
}

@article{somepalli2023understanding,
  title={Understanding and mitigating copying in diffusion models},
  author={Somepalli, Gowthami and Singla, Vasu and Goldblum, Micah and Geiping, Jonas and Goldstein, Tom},
  journal={NeurIPS},
  volume={36},
  year={2023}
}

@inproceedings{wen2024detecting,
  title={Detecting, explaining, and mitigating memorization in diffusion models},
  author={Wen, Yuxin and Liu, Yuchen and Chen, Chen and Lyu, Lingjuan},
  booktitle={ICLR},
  year={2024}
}

@inproceedings{ghosal2023text,
  title={Text-to-audio generation using instruction guided latent diffusion model},
  author={Ghosal, Deepanway and Majumder, Navonil and Mehrish, Ambuj and Poria, Soujanya},
  booktitle={ACM  Multimedia},
  year={2023}
}

@inproceedings{kumari2023ablating,
  title={Ablating concepts in text-to-image diffusion models},
  author={Kumari, Nupur and Zhang, Bingliang and Wang, Sheng-Yu and Shechtman, Eli and Zhang, Richard and Zhu, Jun-Yan},
  booktitle={IEEE/CVF Int. Conf. on Computer Vision},
  year={2023}
}

@inproceedings{sturm2019artificial,
  title={Artificial intelligence and music: open questions of copyright law and engineering praxis},
  author={Sturm, Bob LT and Iglesias, Maria and Ben-Tal, Oded and Miron, Marius and G{\'o}mez, Emilia},
  booktitle={Arts},
  volume={8},
  number={3},
  pages={115},
  year={2019},
  organization={MDPI}
}

@article{fonseca2017freesound,
  title={Freesound datasets: a platform for the creation of open audio datasets},
  author={Fonseca, Eduardo and Pons Puig, Jordi and Favory, Xavier and Font Corbera, Frederic and Bogdanov, Dmitry and Ferraro, Andres and Oramas, Sergio and Porter, Alastair and Serra, Xavier},
  year={2017},
  publisher={ISMIR}
}

@inproceedings{michael_defferrard_2017_1414728,
  author       = {Michaël Defferrard and
                  Kirell Benzi and
                  Pierre Vandergheynst and
                  Xavier Bresson},
  title        = {FMA: A Dataset for Music Analysis.},
  booktitle    = {ISMIR},
  year         = 2017,
  month        = oct,
  venue        = {Suzhou, China},
}

@article{kad,
    author={Chung, Yoonjin and Eu, Pilsun and Lee, Junwon and Choi, Keunwoo and Nam, Juhan and Chon, Ben Sangbae},
    title={KAD: No More FAD! An Effective and Efficient Evaluation Metric for Audio Generation}, 
    journal = {arXiv:2502.15602},
    url = {https://arxiv.org/abs/2502.15602},
    year = {2025}
}

@article{huang2025aligning,
  title={Aligning Text-to-Music Evaluation with Human Preferences},
  author={Huang, Yichen and Novack, Zachary and Saito, Koichi and Shi, Jiatong and Watanabe, Shinji and Mitsufuji, Yuki and Thickstun, John and Donahue, Chris},
  journal={arXiv preprint arXiv:2503.16669},
  year={2025}
}

@article{passoni2025diffused,
  title={Diffused Responsibility: Analyzing the Energy Consumption of Generative Text-to-Audio Diffusion Models},
  author={Passoni, Riccardo and Ronchini, Francesca and Comanducci, Luca and Serizel, Romain and Antonacci, Fabio},
  journal={arXiv preprint arXiv:2505.07615},
  year={2025}
}

@inproceedings{rombach2022high,
  title={High-resolution image synthesis with latent diffusion models},
  author={Rombach, Robin and Blattmann, Andreas and Lorenz, Dominik and Esser, Patrick and Ommer, Bj{\"o}rn},
  booktitle={ IEEE/CVF CVPR},
  year={2022}
}

@inproceedings{majumder2024tango,
  title={Tango 2: Aligning diffusion-based text-to-audio generations through direct preference optimization},
  author={Majumder, Navonil and Hung, Chia-Yu and Ghosal, Deepanway and Hsu, Wei-Ning and Mihalcea, Rada and Poria, Soujanya},
  booktitle={ ACM Multimedia},
  year={2024}
}

@inproceedings{laionclap2023,
  title = {Large-scale Contrastive Language-Audio Pretraining with Feature Fusion and Keyword-to-Caption Augmentation},
  author = {Wu*, Yusong and Chen*, Ke and Zhang*, Tianyu and Hui*, Yuchen and Berg-Kirkpatrick, Taylor and Dubnov, Shlomo},
  booktitle={ICASSP},
  organization={IEEE},
  year = {2023}
}

\end{document}